\documentclass[a4paper,12pt]{article}
\usepackage{amsmath}
\usepackage{amssymb}
\usepackage{amsfonts}
\usepackage{amssymb,amsmath}
\usepackage{cancel}
\usepackage[dvips]{lscape,graphicx,epsfig}
\usepackage{fleqn}
\usepackage[footnotesize]{caption}
\usepackage{mathrsfs}

\def\beq{\begin{equation}}
\def\eeq{\end{equation}}
\def\bea{\begin{eqnarray}}
\def\eea{\end{eqnarray}}

\def\<{\left\langle}
\def\>{\right\rangle}

\newcommand{\bc}{\begin{center}}
\newcommand{\ec}{\end{center}}
\newcommand{\bd}{\begin{displaymath}}
\newcommand{\ed}{\end{displaymath}}
\newcommand{\be}{\begin{equation}}
\newcommand{\ee}{\end{equation}}
\newcommand{\ba}{\begin{array}}
\newcommand{\ea}{\end{array}}
\newcommand{\bt}{\begin{tabular}}
\newcommand{\et}{\end{tabular}}

\newcommand{\ds}{\displaystyle}

\begin{document}

\bibliographystyle{OurBibTeX}

\begin{titlepage}

 \vspace*{-15mm}
\begin{flushright}
\today\\
\end{flushright}
\vspace*{5mm}

\begin{center}
{
\sffamily
\Large
Exotic Higgs decays in the $E_6$ inspired SUSY models
}
\\[8mm]
R.~Nevzorov$^a$\footnote{E-mail: \texttt{nevzorov@itep.ru}},\quad
S.~Pakvasa$^b$\footnote{E-mail: \texttt{pakvasa@phys.hawaii.edu}.}
\\[3mm]
{\small\it
$^a$ Institute for Theoretical and Experimental Physics, \\
Moscow, 117218, Russia \\[2mm]
$^b$ Department of Physics and Astronomy, University of Hawaii,\\
Honolulu, Hawaii 96822, USA }\\[2mm]
\end{center}
\vspace*{0.75cm}

\begin{abstract}
\noindent
We study the decays of the SM-like Higgs state within the $E_6$ inspired
supersymmetric (SUSY) models with exact custodial symmetry that forbids
tree-level flavor-changing transitions and the most dangerous baryon and
lepton number violating operators. In these models there are two states
which are absolutely stable and can contribute to the dark matter density.
One of them is the lightest SUSY particle (LSP) which is expected to be
lighter than $1\,\mbox{eV}$ forming hot dark matter in the Universe. The
presence of another stable neutral state allows to account for the observed
cold dark matter density. In the considered SUSY models next--to--lightest
SUSY particle (NLSP) also tend to be light. We argue that the NLSP with GeV
scale mass can result in the substantial branching ratio of the nonstandard
decays of the lightest Higgs boson.
\end{abstract}

\end{titlepage}
\newpage
\setcounter{footnote}{0}

\section{Introduction}

The observation of a new bosonic state with a mass around $\sim
125\mbox{GeV}$ \cite{:2012gk,:2012gu} may provide a window into
new physics beyond the Standard Model (SM). At the moment the
observed signal strengths are consistent with the SM Higgs boson
but more data is needed to assess the nature of the recently
discovered state. Physics beyond the SM may affect the Higgs decay
rates to SM particles and give rise to new channels of Higgs
decays (for recent reviews of nonstandard Higgs boson decays see
\cite{Chang:2008cw}). In particular, Higgs boson can decay with a
substantial branching fraction into states which can not be
directly detected. Such invisible Higgs decay modes may occur in
models with an enlarged symmetry breaking sector (Majoron models,
SM with extra singlet scalar fields etc.) \cite{majoron,
Martin:1999qf}, in ``hidden valley'' models \cite{hidden-valley},
in the SM with a fourth generation of fermions
\cite{fourth-generation}, in the supersymmetric (SUSY) extensions
of the SM \cite{Baer:1987eb}\footnote{Recently the nonstandard
Higgs decays within the Next--to--Minimal Supersymmetric Standard
model were discussed in \cite{King:2012tr}.}, in the models with
compact and large extra dimensions \cite{Martin:1999qf, higgs-extraD},
 in the littlest Higgs model with T-parity \cite{Asano:2006nr} etc.

In the context of invisible Higgs decays it is especially interesting to consider
the nature and extent of invisibility acquired by the SM--like Higgs state within
well motivated SUSY extensions of the SM. Here we focus on the $E_6$ inspired SUSY
models which are based on the low--energy SM gauge group together with an extra
$U(1)_{N}$ gauge symmetry defined by:
\be
U(1)_N=\ds\frac{1}{4} U(1)_{\chi}+\ds\frac{\sqrt{15}}{4} U(1)_{\psi}\,.
\label{1}
\ee
The\quad two\quad anomaly-free\quad $U(1)_{\psi}$\quad and $U(1)_{\chi}$
symmetries can originate from the breakings
$E_6\to$ $SO(10)\times U(1)_{\psi}$, $SO(10)\to SU(5)\times U(1)_{\chi}$.
To ensure anomaly cancellation the particle spectrum in these models
is extended to fill out three complete 27-dimensional representations
of the gauge group $E_6$. Each $27$-plet contains one generation of ordinary matter;
singlet fields, $S_i$; up and down type Higgs doublets, $H^{u}_{i}$ and $H^{d}_{i}$;
charged $\pm 1/3$ coloured exotics $D_i$, $\bar{D}_i$. The presence of exotic
matter in $E_6$ inspired SUSY models generically lead to non--diagonal flavour
transitions and rapid proton decay. To suppress flavour changing processes as well
as baryon and lepton number violating operators one can impose a set of discrete
symmetries \cite{King:2005jy}--\cite{King:2005my}.
The $E_6$ inspired SUSY models with extra $U(1)_{N}$ gauge symmetry and
suppressed flavor-changing transitions, as well as baryon number violating
operators allow exotic matter to survive down to the TeV scale that may lead
to spectacular new physics signals at the LHC which were analysed in
\cite{King:2005jy}--\cite{Accomando:2006ga}. Only in this
Exceptional Supersymmetric Standard Model (E$_6$SSM) \cite{King:2005jy}--\cite{King:2005my}
right--handed neutrinos do not participate in the gauge interactions so that
they may be superheavy, shedding light on the origin of the mass hierarchy in the
lepton sector and providing a mechanism for the generation of the baryon asymmetry
in the Universe via leptogenesis \cite{King:2008qb}.
Recently the particle spectrum and collider signatures associated with it were
studied within the constrained version of the E$_6$SSM \cite{8}.

In this note we consider the nonstandard Higgs decays within the $E_6$ inspired SUSY
models in which a single discrete $\tilde{Z}^{H}_2$ symmetry forbids tree-level flavor-changing
transitions and the most dangerous baryon and lepton number violating operators \cite{nevzorov}.
These models contain at least two states which are absolutely stable and can contribute to the
relic density of dark matter. One of these states is a lightest SUSY particle
(LSP) while another one tends to be the lightest ordinary neutralino. The masses of
the LSP and next--to--lightest SUSY particle (NLSP) are determined by the vacuum
expectation values (VEVs) of the Higgs doublets. As a consequence they give rise
to nonstandard Higgs boson decays. In the phenomenologically viable scenarios LSP
should have mass around $1\,\mbox{eV}$ or even smaller forming hot dark matter
in the Universe while NLSP can be substantially heavier. NLSPs with GeV scale
masses result in substantial branching ratios of the lightest Higgs decays into
NLSPs. Since NLSP tend to be longlived particle in this case it decays outside
the detectors leading to the invisible decays of the SM-like Higgs state. In the
considered $E_6$ inspired SUSY model the lightest ordinary neutralino can account
for all or some of the observed cold dark matter relic density.

The paper is organised as follows. In the next section we briefly review
the $E_6$ inspired SUSY models with exact custodial $\tilde{Z}^{H}_2$ symmetry.
In section 3 we specify a set of benchmark scenarios that lead to the invisible decays
of the SM--like Higgs state mentioned above. Section 4 concludes the paper.

\section{$E_6$ inspired SUSY models with exact $\tilde{Z}^{H}_2$ symmetry}

In this section, we give a brief review of the $E_6$ inspired SUSY models with
exact custodial $\tilde{Z}^{H}_2$ symmetry \cite{nevzorov}. These models imply that
near some high energy scale (scale $M_X$) $E_6$ or its subgroup is broken down to
$SU(3)_C\times SU(2)_W\times U(1)_Y\times U(1)_{\psi}\times U(1)_{\chi}\times Z_{2}^{M}$,
where $Z_{2}^{M}=(-1)^{3(B-L)}$ is a matter parity. Below scale $M_X$ the particle content
of the considered models involves three copies of $27_i$--plets and a set of $M_{l}$ and
$\overline{M}_l$ supermultiplets from the incomplete $27'_l$ and $\overline{27'}_l$
representations of $E_6$. All matter superfields, that fill in complete $27_i$--plets, are odd
under $\tilde{Z}^{H}_2$ discrete symmetry while the supermultiplets $\overline{M}_l$ can
be either odd or even. All supermultiplets $M_{l}$ are even under the $\tilde{Z}^{H}_2$
symmetry and therefore can be used for the breakdown of gauge symmetry. In the simplest
case the set of $M_{l}$ includes $H_u$, $H_d$, $S$ and $L_4$, where $L_4$ and
$\overline{L}_4$ are lepton $SU(2)_W$ doublet and anti--doublet supermultiplets that
originate from a pair of additional $27'_{L}$ and $\overline{27'}_L$.

At low energies (i.e. TeV scale) the superfields $H_u$, $H_d$ and $S$ play the role of
Higgs fields. The VEVs of these superfields ($\langle H_d \rangle = v_1/\sqrt{2}$,
$\langle H_u \rangle = v_2/\sqrt{2}$ and $\langle S \rangle = s/\sqrt{2}$) break the
$SU(2)_W\times U(1)_Y\times U(1)_{N}$ gauge symmetry down to $U(1)_{em}$ associated
with the electromagnetism. In the simplest scenario $\overline{H}_u$, $\overline{H}_d$
and $\overline{S}$ are odd under the $\tilde{Z}^{H}_2$ symmetry. As a consequence
$\overline{H}_u$, $\overline{H}_d$ and $\overline{S}$ from the $\overline{27'}_l$
get combined with the superposition of the corresponding components from $27_i$ so that the
resulting vectorlike states gain masses of order of $M_X$. On the other hand $L_4$ and
$\overline{L}_4$ are even under the $\tilde{Z}^{H}_2$ symmetry. These supermultiplets
form TeV scale vectorlike states to render the lightest exotic quark unstable. In this
simplest scenario the exotic quarks are leptoquarks.

The $\tilde{Z}^{H}_2$ symmetry allows the Yukawa interactions in the superpotential that
originate from $27'_l \times 27'_m \times 27'_n$ and $27'_l \times 27_i \times 27_k$.
One can easily check that the corresponding set of operators does not contain any
that lead to the rapid proton decay. Since the set of multiplets $M_{l}$ contains only one
pair of doublets $H_d$ and $H_u$ the $\tilde{Z}^{H}_2$ symmetry also forbids unwanted
FCNC processes at the tree level. The gauge group and field content of the $E_6$ inspired
SUSY models considered here can originate from the orbifold GUT models in which the splitting
of GUT multiplets can be naturally achieved \cite{nevzorov}.

In the simplest scenario discussed above extra matter beyond the minimal supersymmetric
standard model (MSSM) fill in complete $SU(5)$ representations. As a result the gauge
coupling unification remains almost exact in the one--loop approximation. It was also
shown that in the two--loop approximation the unification of the gauge couplings in the
considered scenario can be achieved for any phenomenologically acceptable value of
$\alpha_3(M_Z)$, consistent with the central measured low energy value \cite{unif-e6ssm}.

\begin{table}[ht]
\centering
\begin{tabular}{|c|c|c|c|c|c|c|c|}
\hline
                   &  $27_i$          &   $27_i$              &$27'_{H_u}$&$27'_{S}$&
$\overline{27'}_{H_u}$&$\overline{27'}_{S}$&$27'_{L}$\\
& & &$(27'_{H_d})$& &$(\overline{27'}_{H_d})$& &$(\overline{27'}_L)$\\
\hline
                   &$Q_i,u^c_i,d^c_i,$&$\overline{D}_i,D_i,$  & $H_u$     & $S$     &
$\overline{H}_u$&$\overline{S}$&$L_4$\\
                   &$L_i,e^c_i,N^c_i$ &  $H^d_{i},H^u_{i},S_i$& $(H_d)$   &         &
$(\overline{H}_d)$& &$(\overline{L}_4)$\\
\hline
$\tilde{Z}^{H}_2$  & $-$              & $-$                   & $+$       & $+$     &
$-$&$-$&$+$\\
\hline
$Z_{2}^{M}$        & $-$              & $+$                   & $+$       & $+$     &
$+$&$+$&$-$\\
\hline
$Z_{2}^{E}$        & $+$              & $-$                   & $+$       & $+$     &
$-$&$-$&$-$\\
\hline
\end{tabular}
\caption{Transformation properties of different components of $E_6$ multiplets
under $\tilde{Z}^H_2$, $Z_{2}^{M}$ and $Z_{2}^{E}$ discrete symmetries.}
\label{tab1}
\end{table}

As mentioned before, the gauge symmetry in the $E_6$ inspired SUSY models being
considered here, is broken so that the low--energy effective Lagrangian of these models is
invariant under both $Z_{2}^{M}$ and $\tilde{Z}^{H}_2$ symmetries. Since
$\tilde{Z}^{H}_2 = Z_{2}^{M}\times Z_{2}^{E}$  the $Z_{2}^{E}$ symmetry associated
with exotic states is also conserved. The transformation properties of different
components of $27_i$, $27'_l$ and $\overline{27'}_l$ supermultiplets under the
$\tilde{Z}^{H}_2$, $Z_{2}^{M}$ and $Z_{2}^{E}$ symmetries are summarized in
Table~\ref{tab1}. The invariance of the Lagrangian under the $Z_{2}^{E}$ symmetry
implies that the lightest exotic state, which is odd under this symmetry, must be
stable. Using the method proposed in \cite{Hesselbach:2007te} it was argued that
that there are theoretical upper bounds on the masses of the lightest and second
lightest inert neutralino states \cite{Hall:2010ix}\footnote{We use the terminology
``Inert Higgs'' to denote Higgs--like doublets and SM singlets that do not develop VEVs.
The fermionic components of these supermultiplets form inert neutralino and chargino
states.}. These states are predominantly the fermion components of the two SM singlet
superfields $S_i$ from $27_i$ which are odd under the $Z_{2}^{E}$ symmetry.
Their masses do not exceed $60-65\,\mbox{GeV}$ so that the lightest and second
lightest inert neutralino states ($\tilde{H}^0_1$ and $\tilde{H}^0_2$) tend to be the
lightest exotic particles in the spectrum \cite{Hall:2010ix}.

The $Z_{2}^{M}$ symmetry conservation ensures that $R$--parity is also conserved.
Since the lightest inert neutralino $\tilde{H}^0_1$ is also the lightest $R$--parity
odd state either the lightest $R$--parity even exotic state or the lightest $R$--parity
odd state with $Z_{2}^{E}=+1$ must be absolutely stable. In the considered $E_6$
inspired SUSY models most commonly the second stable state is the lightest ordinary
neutralino $\chi_1^0$ ($Z_{2}^{E}=+1$). Both stable states are natural dark matter
candidates.

When $|m_{\tilde{H}^0_{1}}|\ll M_Z$ the couplings of the lightest inert neutralino
to the gauge bosons, Higgs states, quarks and leptons are very small resulting in
very small annihilation cross section for $\tilde{H}^0_1\tilde{H}^0_1\to \mbox{SM particles}$,
making the cold dark matter density much larger than its measured value. In principle,
$\tilde{H}^0_1$ could account for all or some of the observed cold dark matter density if it
had a mass close to half the $Z$ mass. In this case the lightest inert neutralino states
annihilate mainly through an $s$--channel $Z$--boson \cite{Hall:2010ix}, \cite{Hall:2009aj}.
However the usual SM-like Higgs boson decays more than 95\% of the time into either $\tilde{H}^0_1$
or $\tilde{H}^0_2$ in these cases while the total branching ratio into SM particles is suppressed.
Because of this the corresponding scenarios are basically ruled out nowadays \cite{Hall:2010ix}.

The simplest phenomenologically viable scenarios imply that the lightest inert neutralinos
are extremely light. For example, these states can be substantially lighter than
$1\,\mbox{eV}$\footnote{The presence of very light neutral fermions in the particle spectrum
might have interesting implications for the neutrino physics (see, for example \cite{Frere:1996gb}).}.
In this case, light $\tilde{H}^0_1$ forms hot dark matter in the Universe but gives only a very minor contribution
to the dark matter density while the lightest ordinary neutralino may account for all or some of
the observed cold dark matter density.

\section{Nonstandard Higgs decays}

As discussed earlier, the $E_6$ inspired SUSY models considered here involves three families of
up and down type Higgs--like doublet supermultiplets ($H^{u}_{i}$ and $H^{d}_{i}$) and three
SM singlet superfields ($S_i$) that carry $U(1)_{N}$ charges. One family of the Higgs--like
doublets and one SM singlet develop VEVs breaking gauge symmetry. The fermionic components of
other Higgs--like and singlet superfields form inert neutralino and chargino states.
The Yukawa interactions of inert Higgs superfields are described by the superpotential
\begin{eqnarray}
W_{IH}=\lambda_{\alpha\beta} S (H^d_{\alpha} H^u_{\beta})+
f_{\alpha\beta} S_{\alpha} (H_d H^u_{\beta})+
\tilde{f}_{\alpha\beta} S_{\alpha} (H^d_{\beta} H_u)\,,
\label{2}
\end{eqnarray}
where $\alpha,\beta=1,2$\,. Without loss of generality it is always possible to choose the basis
so that $\lambda_{\alpha\beta}=\lambda_{\alpha\alpha}\,\delta_{\alpha\beta}$. In this basis the
masses of inert charginos are given by
\begin{equation}
m_{\tilde{H}^{\pm}_{\alpha}}=\dfrac{\lambda_{\alpha\alpha}}{\sqrt{2}}\,s\,.
\label{3}
\end{equation}

In our analysis here we shall choose the VEV of the SM singlet field $s$
to be large enough ($s\simeq 12\,\mbox{TeV}$) to ensure that the experimental
constraints on $Z'$ boson mass ($M_{Z'}\gtrsim 2\,\mbox{TeV}$) and $Z-Z'$ mixing
are satisfied. To avoid the LEP lower limit on the masses of inert charginos
we also choose the Yukawa couplings $\lambda_{\alpha\alpha}$ so that all inert
chargino states have masses which are larger than $100\,\mbox{GeV}$.
In the following analysis we also require the validity of perturbation theory up
to the GUT scale that constrains the allowed range of all Yukawa couplings.

Here we restrict our attention to the part of the parameter space that corresponds
to $\lambda_{\alpha\alpha} s\gg f_{\alpha\beta} v,\, \tilde{f}_{\alpha\beta} v$.
In that  limit the inert neutralino states which are predominantly
linear superpositions of the neutral components of inert Higgsinos, i.e. $\tilde{H}^{d0}_{\alpha}$
and $\tilde{H}^{u0}_{\alpha}$, are normally heavier than $100\,\mbox{GeV}$ and can be integrated
out. Then the resulting $2\times 2$ mass matrix can be written as follows
\begin{equation}
\begin{array}{c}
M_{IS}=
-\dfrac{v^2\sin 2\beta}{4 m_{\tilde{H}^{\pm}_1}}
\left(
\begin{array}{cc}
2\tilde{f}_{11} f_{11}                         & \tilde{f}_{11} f_{21} + f_{11} \tilde{f}_{21}\\[2mm]
\tilde{f}_{11} f_{21} + f_{11} \tilde{f}_{21}  & 2\tilde{f}_{21} f_{21}
\end{array}
\right)\qquad\qquad\qquad\qquad\qquad\\
\qquad\qquad\qquad\qquad\qquad-
\dfrac{v^2\sin 2\beta}{4 m_{\tilde{H}^{\pm}_2}}
\left(
\begin{array}{cc}
2\tilde{f}_{12} f_{12}                         & \tilde{f}_{12} f_{22} + f_{12} \tilde{f}_{22}\\[2mm]
\tilde{f}_{12} f_{22} + f_{12} \tilde{f}_{22}  & 2\tilde{f}_{22} f_{22}
\end{array}
\right)
\,,
\end{array}
\label{4}
\end{equation}
where $v=\sqrt{v_1^2+v_2^2}\simeq 246\,\mbox{GeV}$ and $\tan\beta=v_2/v_1$.
The mass matrix (\ref{4}) can be easily diagonalized. Two lightest inert neutralino states $\tilde{H}^0_{1}$
and $\tilde{H}^0_{2}$ are predominantly inert singlinos. In our limit these states tend to be
substantially lighter than $100\,\mbox{GeV}$.

When the SUSY breaking scale $M_S$ is considerably larger than the electroweak (EW) scale, the mass matrix
of the CP--even Higgs sector has a hierarchical structure and can be also diagonalized using the perturbation
theory \cite{Nevzorov:2001um}--\cite{Nevzorov:2004ge}. Here we are going to focus on the scenarios with
moderate values of $\tan\beta$ ($\tan\beta< 2-3$). For these values of $\tan\beta$ the mass of the lightest
CP--even Higgs boson $m_{h_1}$ is very sensitive to the choice of the coupling $\lambda(M_t)$. In particular,
in order to get $m_{h_1}\simeq 125\,\mbox{GeV}$ the coupling $\lambda(M_t)$ must be larger than $g'_1\simeq 0.47$.
When $\lambda\gtrsim g'_1$, the qualitative pattern of the Higgs spectrum is rather similar
to the one which arises in the PQ symmetric NMSSM \cite{Miller:2005qua}--\cite{Nevzorov:2004ge}.
In the considered limit the heaviest CP--even, CP--odd and charged states are almost degenerate and lie beyond
the $\mbox{TeV}$ range while the mass of the second lightest CP--even Higgs state is set by $M_{Z'}$
\cite{King:2005jy}. In this case the lightest CP--even Higgs boson is the analogue of the SM Higgs field.

The lightest and second lightest inert neutralinos interact with the $Z$--boson and the SM--like Higgs state.
The corresponding part of the Lagrangian, that describes these interactions, can be presented in the following form:
\begin{equation}
\begin{array}{c}
\mathcal{L}_{Zh}=\sum_{\alpha,\beta}\dfrac{M_Z}{2 v}Z_{\mu}
\biggl(\tilde{H}^{0T}_{\alpha}\gamma_{\mu}\gamma_{5}\tilde{H}^0_{\beta}\biggr) R_{Z\alpha\beta}
\qquad\qquad\qquad\qquad\qquad\qquad\\
\qquad\qquad\qquad\qquad\qquad\qquad+
\sum_{\alpha,\beta} (-1)^{\theta_{\alpha}+\theta_{\beta}} X^{h}_{\alpha\beta} \biggl(\psi^{0T}_{\alpha}
(-i\gamma_{5})^{\theta_{\alpha}+\theta_{\beta}}\psi^0_{\beta}\biggr) h\,,\\[4mm]
\end{array}
\label{5}
\end{equation}
where $\alpha,\beta=1,2$. In Eq.~(\ref{5}) $\psi^0_{\alpha}=(-i\gamma_5)^{\theta_{\alpha}}\tilde{H}^0_{\alpha}$ is
the set of inert neutralino eigenstates with positive eigenvalues, while $\theta_{\alpha}$ equals 0 (1) if the eigenvalue
corresponding to $\tilde{H}^0_{\alpha}$ is positive (negative). The inert neutralinos are labeled according to increasing
absolute value of mass, with $\tilde{H}^0_1$ being the lightest inert neutralino.

We further assume that the lightest inert neutralino is substantially lighter than $1\,\mbox{eV}$ so that it gives
only a very minor contribution to the dark matter density. On the other hand we allow the second lightest inert neutralino
state to have mass in the GeV range. Although these states are substantially lighter than $100\,\mbox{GeV}$ their couplings
to the $Z$--boson can be rather small because of the inert singlino admixture in these states. Therefore any possible signal
which these neutralinos could give rise to at former colliders would be extremely suppressed and such states could remain
undetected.

The couplings of the Higgs states to the inert neutralinos originate from the superpotential (\ref{2}). If all Higgs states
except the lightest one are much heavier than the EW scale then the couplings of the SM--like Higgs boson to the lightest
and second lightest inert neutralinos are determined by their masses \cite{Hall:2010ix}. Since we assumed that the mass
of $\tilde{H}^0_1$ is lighter than $1\,\mbox{eV}$ the couplings of the lightest Higgs boson to $\tilde{H}^0_1\tilde{H}^0_1$
and $\tilde{H}^0_1\tilde{H}^0_2$ are negligibly small and can be ignored in our analysis. Also because of this the experiments
for the direct detection of dark matter do not set any stringent constraints on the masses and couplings of the lightest and
second lightest inert neutralinos. In the considered case the coupling of the SM--like Higgs state to $\tilde{H}^0_2$
is basically proportional to the second lightest inert neutralino mass divided by the VEV, i.e.
$X^{h}_{22}\simeq |m_{\tilde{H}^0_{2}}|/v$ \cite{Hall:2010ix}. This coupling gives rise to the decays of the lightest
Higgs boson into $\tilde{H}^0_2$ pairs with partial widths given by
\be
\Gamma(h_1\to\tilde{H}^0_{2}\tilde{H}^0_{2})=\dfrac{(X^{h}_{22})^2 m_{h_1}}{4\pi}\biggl(
1-4\dfrac{|m_{\tilde{H}^0_{2}}|^2}{m^2_{h_1}}\biggr)^{3/2}\,.
\label{6}
\ee

In order to compare the partial widths associated with the exotic decays of the SM-like Higgs state (\ref{6})
with the Higgs decay rates into the SM particles we shall specify a set of benchmark points (see Table 2).
For each benchmark scenario we calculate the spectrum of the inert neutralinos, inert charginos
and Higgs bosons as well as their couplings and the branching ratios of the nonstandard decays of the lightest
CP-even Higgs state. We fix $\tan\beta=1.5$ and $\lambda(M_t)=0.6$. As it was mentioned before, such a large
value of $\lambda(M_t)$ allows $m_{h_1}$ to be $125\,\mbox{GeV}$ for moderate $\tan\beta$.
In addition, we set stop scalar masses to be equal to $m_Q=m_U=M_S=4\,\mbox{TeV}$ and restrict our consideration
to the so-called maximal mixing scenario when the stop mixing parameter $X_t=A_t-\lambda s/(\sqrt{2}\tan\beta)$
is equal to $X_t=\sqrt{6} M_S$. From Table 2 it follows that the structure of the Higgs spectrum is extremely
hierarchical. In Table 2 the masses of the heavy Higgs states are computed in the leading one--loop approximation.
In the case of the lightest Higgs boson mass the leading two--loop corrections are taken into account.

Since the structure of the Higgs spectrum is very hierarchical, the partial decay widths that correspond
to the decays of the lightest CP-even Higgs state into the SM particles are basically the same as in the SM.
Because of this, for the calculation of the Higgs decay rates into the SM particles we use the results presented
in \cite{King:2012is} where these rates were computed within the SM for different values of the Higgs mass.
When $m_{h_1}\simeq 125\,\mbox{GeV}$ the SM-like Higgs state decays predominantly into $b$-quark.
In the SM the corresponding branching ratio is about $60\%$ whereas the branching ratios associated with
Higgs decays into $WW$, $ZZ$ and $\gamma\gamma$ are about $20\%$, $2.1\%$ and $0.23\%$ respectively \cite{King:2012is}.
The total decay width of the Higgs boson near 125 GeV is $3.95\,\mbox{MeV}$.

For the calculation of the Higgs decay rates into $\tilde{H}^0_2\tilde{H}^0_2$ we use Eq.~(\ref{6}).
From this equation one can see that the branching ratios of the SM--like Higgs state into the second lightest inert
neutralinos depend rather strongly on the masses of these exotic particles. When $\tilde{H}^0_2$ is relatively heavy,
i.e. $m_{\tilde{H}^0_{2}} \gg m_b(m_{h_1})$, the lightest Higgs boson decays predominantly into $\tilde{H}^0_2\tilde{H}^0_2$
while the branching ratios for decays into SM particles are suppressed. To ensure that the observed signal associated
with the Higgs decays into $\gamma\gamma$ is not too much suppressed we restrict our consideration here to the
GeV scale masses of the second lightest inert neutralino.

The set of the benchmark points (i)-(iv) that we specify in Table 2 demonstrates that one can get extremely light
$\tilde{H}^0_1$ with mass $\sim 0.1-0.01\,\mbox{eV}$, relatively light $\tilde{H}^0_2$, that has a mass of the order
of $1-0.1\,\mbox{GeV}$, and a relatively small value of the coupling $R_{Z12}$ that allows the second lightest inert
neutralino to decay within a reasonable time. In these benchmark scenarios the second lightest inert neutralino
decays into the lightest one and a fermion--antifermion pair via virtual $Z$. Since $R_{Z12}$ is relatively
small $\tilde{H}^0_2$ tend to have a long lifetime. If the second lightest inert neutralino state decays during or
after Big Bang Nucleosynthesis (BBN) it may destroy the agreement between the predicted and observed light element
abundances. To preserve the success of the BBN, $\tilde{H}^0_2$ should decay before BBN, i.e. its lifetime $\tau_{\tilde{H}^0_{2}}$
has to be smaller than something like $1\,\mbox{sec}.$ This requirement constrains $|R_{Z12}|$. Indeed, for
$m_{\tilde{H}^0_{2}}=1\,\mbox{GeV}$ the absolute value of the coupling $R_{Z12}$ should be larger than
$1\cdot 10^{-6}$ \cite{King:2012wg}. On the other hand the value of $|R_{Z12}|$ becomes smaller when
the mass of the lightest inert neutralino decreases. Therefore in general sufficiently large fine tuning is needed to 
ensure that $|R_{Z12}| \gtrsim 10^{-6}$ for sub-eV lightest inert neutralino state. 
The constraint on $|R_{Z12}|$ becomes much more stringent with decreasing 
$m_{\tilde{H}^0_{2}}$ because $\tau_{\tilde{H}^0_{2}}\sim 1/(|R_{Z12}|^2 m_{\tilde{H}^0_{2}}^5)$. As a result, it
is somewhat problematic to satisfy this restriction for $m_{\tilde{H}^0_{2}}\lesssim 100\,\mbox{MeV}$.

The benchmark scenarios (i)-(iv) presented in Table 2 indicate that the branching ratio of the decays of SM--like
Higgs boson into second lightest inert neutralino can vary from $0.2\%$ to $20\%$ (i.e. from $0\%$ to $20\%$
for practical purposes) when $m_{\tilde{H}^0_{2}}$ changes from $0.3\,\mbox{GeV}$ to $2.7\,\mbox{GeV}$.
For smaller (larger) values of the second lightest inert neutralino masses, the branching ratio associated with these
nonstandard decays of the lightest CP--even Higgs states is even smaller (larger). At the same time the couplings of
$\tilde{H}^0_1$ and $\tilde{H}^0_2$ to the $Z$--boson are so tiny that the lightest and second lightest inert neutralino states
could not be observed before. In particular, their contribution to the $Z$--boson width tend to be rather small. The $Z$--boson
invisible width is characterized by the effective number of neutrino species $N_{\nu}^{eff}$. Its measured value is
$N_{\nu}^{exp}=2.984\pm 0.008$ \cite{pdg} whereas in the SM $N_{\nu}^{eff}=3$. The contributions of the lightest
and second lightest inert neutralino states to the $Z$--boson width can be parameterized similarly. In the case of benchmark
scenarios (i), (ii), (iii), (iv) the effective numbers of neutrino species associated with these contributions are $5.8\cdot 10^{-5}$,
$0.002$, $3.7\cdot 10^{-6}$ and 0.011 respectively.

The second lightest inert neutralino states, that originate from the decays of the SM--like Higgs boson, sequentially decay into
$\tilde{H}^0_1$ and pairs of leptons and quarks via virtual $Z$. Thus, in principle, the exotic decays of the lightest CP--even
Higgs state results in two fermion--antifermion pairs and missing energy in the final state. Nevertheless because coupling
$R_{Z12}$ is quite small $\tilde{H}^0_2$ tend to live longer than $10^{-8}\,\mbox{sec}$. As a consequence the
second lightest inert neutralino state typically decays outside the detectors and will not be observed at the LHC. Therefore the
decay channel $h_1\to\tilde{H}^0_2\tilde{H}^0_2$ normally give rise to an invisible branching ratio of the SM--like Higgs boson.
Such invisible decays of the lightest CP--even Higgs state take place in the benchmark scenarios (i), (iii) and (iv). In the case
of benchmark scenario (ii) the absolute value of $R_{Z12}$ coupling is larger than in other benchmark scenarios so that
$\tau_{\tilde{H}^0_{2}}\sim 10^{-11}\,\mbox{sec}$. In this case some of the decay products of $\tilde{H}^0_2$ produced through
the decays $h_1\to\tilde{H}^0_2\tilde{H}^0_2$ might be observed at the LHC. In particular, we hope that it might be possible to
detect the relatively energetic $\mu^{+} \mu^{-}$ pairs that come from these exotic decays of the lightest CP--even Higgs state.

\section{Conclusions}

In this note we have considered the nonstandard decays of the lightest Higgs boson within well motivated
SUSY extensions of the SM based on the $SU(3)_C\times SU(2)_W\times U(1)_Y\times U(1)_{N}\times Z_{2}^{M}$
symmetry. The low energy matter content of these $E_6$ inspired models includes three $27$ representations of $E_6$
and a pair of $SU(2)_W$ doublets $L_4$ and $\overline{L}_4$. In particular, the low--energy spectrum of the SUSY models
being considered here involves three families of Higgs--like doublets $H^{d}_{i}$ and $H^{u}_{i}$, three families of
exotic quarks $D_i$ and $\bar{D}_i$ as well as three SM singlets $S_i$ that carry $U(1)_{N}$ charges. In order to suppress
flavour changing processes at the tree--level and forbid the most dangerous baryon and lepton number violating operators
we imposed $\tilde{Z}^{H}_2$ discrete symmetry under which one pair of Higgs--like doublet supermultiplets, one SM-type
singlet superfield, $L_4$ and $\overline{L}_4$ are even while all other superfields are odd. The pair of the Higgs--like doublets
and SM singlet, which are even under $\tilde{Z}^{H}_2$ symmetry, acquire VEVs forming a Higgs sector. The fermionic
components of the Higgs--like and SM singlet superfields, which are $\tilde{Z}^{H}_2$ odd, compose a set of inert neutralino
and chargino states. The lightest and second lightest inert neutralino ($\tilde{H}^0_1$ and $\tilde{H}^0_2$) which are
predominantly inert singlinos tend to be LSP and NLSP in these $E_6$ inspired SUSY models. In the simplest phenomenologically
viable scenarios LSP is expected to be substantially lighter than $1\,\mbox{eV}$ and form hot dark matter in the Universe. Since
LSP is so light it gives only minor contribution to the dark matter density. Because of the conservation of the $Z^{M}_2$
and $\tilde{Z}^{H}_2$ symmetries the lightest ordinary neutralino can be also absolutely stable and may account for
all or some of the observed cold dark matter density.

The presence of light LSP and NLSP in the particle spectrum gives rise to new decay channels of the $Z$--boson and
the SM--like Higgs state. In order to illustrate this, we specified a set of the benchmark points. The results of
our analysis indicate that the couplings of $\tilde{H}^0_1$ and $\tilde{H}^0_2$ to the $Z$--boson can be very
small so that these states could escape detection at former and present experiments. The couplings of the SM--like
Higgs boson to the LSP and NLSP are determined by their masses. Since $\tilde{H}^0_1$ is expected to be extremely light it
does not affect Higgs phenomenology. At the same time we argued that the NLSP with the GeV scale masses gives rise to the
substantial branching ratio of the nonstandard decays of the lightest Higgs boson, i.e $h_1\to \tilde{H}^0_2\tilde{H}^0_2$.
After being produced the second lightest inert neutralino states sequentially decay into the LSP and pairs of leptons and
quarks via virtual $Z$. Thus these decays of the lightest CP--even Higgs state lead to two fermion--antifermion pairs and
missing energy in the final state. However due to the small couplings of the NLSP to the $Z$--boson the second lightest
inert neutralino states tend to decay outside the detectors resulting in the invisible branching ratio of the lightest
CP--even Higgs boson.

The branching fraction of the nonstandard Higgs decays depend rather strongly on the mass of the NLSP.
When $m_{\tilde{H}^0_{2}} \gg m_b(m_{h_1})$ the lightest Higgs boson decays mainly into $\tilde{H}^0_2\tilde{H}^0_2$
leading to the suppression of the branching ratios for the decays of $h_1$ into SM particles. To avoid such suppression
we restrict our consideration to the GeV scale masses of $\tilde{H}^0_2$. On the other hand we also required that the
second lightest inert neutralino states decay before BBN, i.e. their lifetime is shorter than $1\,\mbox{sec}$. This
requirement rules out too light $\tilde{H}^0_2$ because $\tau_{\tilde{H}^0_{2}}\sim 1/(m_{\tilde{H}^0_{2}}^5)$.
Our numerical analysis indicates that it is rather problematic to satisfy this restriction for
$m_{\tilde{H}^0_{2}}\lesssim 100\,\mbox{MeV}$.
The set of the benchmark points that we specified demonstrates that the branching ratio of the non-standard decays of
the lightest Higgs state can be as large as 20-30\% if the second lightest inert neutralino is heavier than 2.5 GeV.
When this inert neutralino state is lighter than 0.5 GeV the corresponding branching ratio is suppressed so that
it can be as small as $10^{-3}-10^{-4}$.

\section*{Acknowledgements}
\vspace{-2mm} We would like to thank S.~F.~King, M.~Sher and X.~Tata for  fruitful discussions.
R.N. is also grateful to E.~Boos, M.~Dubinin, S.~Demidov, D.~Gorbunov, M.~Libanov,
V.~Novikov, O.~Kancheli, D.~Kazakov,  M.~Muhlleitner, V.~Rubakov, S.~Troitsky, M.~Vysotsky for valuable comments
and remarks. One of us (S.P.) was supported in part by the U.S. D.O.E. Grant DE-FG02-04ER41291 and
would also like to acknowledge the support of the Alexander von Humboldt Foundation with a Research
Award and the hospitality of Professor Heinrich Paes and the Physics Department of the Technical
University of Dortmund while this work was being completed.

\begin{table}[ht]
\centering
\begin{tabular}{|c||c|c|c|c|}
\hline
                              & i       &   ii      &   iii     &   iv    \\\hline\hline
$\lambda_{22}$                & -0.03   &   -0.012  &   -0.06   &   0  \\\hline
$\lambda_{21}$                & 0       &   0       &   0       &   0.02   \\\hline
$\lambda_{12}$                & 0       &   0       &   0       &   0.02   \\\hline
$\lambda_{11}$                & 0.03    &   0.012   &   0.06    &   0  \\\hline\hline
$f_{22}$                      & -0.1    &   -0.1    &   -0.1    &   0.6    \\\hline
$f_{21}$                      & -0.1    &   -0.1    &   -0.1    &   0.00245 \\\hline
$f_{12}$                      & 0.00001 &   0.00001 &   0.00001 &   0.00245 \\\hline
$f_{11}$                      & 0.1     &   0.1     &   0.1     &   0.00001 \\\hline\hline
$\tilde{f}_{22}$              & 0.1     &   0.1     &   0.1     &   0.6     \\\hline
$\tilde{f}_{21}$              & 0.1     &   0.1     &   0.1     &   0.002 \\\hline
$\tilde{f}_{12}$              & 0.000011&   0.000011&   0.000011&   0.002  \\\hline
$\tilde{f}_{11}$              & 0.1     &   0.1     &   0.1     &   0.00001 \\\hline\hline
$|m_{\tilde{\chi}^0_1}|$/GeV  &$2.7\cdot 10^{-11}$  &   $6.5\cdot 10^{-11}$ & $1.4\cdot 10^{-11}$   & $0.31\cdot 10^{-9}$ \\\hline
$|m_{\tilde{\chi}^0_2}|$/GeV  & 1.09    &   2.67    &   0.55    &   0.319   \\\hline
$|m_{\tilde{\chi}^0_3}|$/GeV  & 254.6   &   101.8   &   509.1   &   169.7  \\\hline
$|m_{\tilde{\chi}^0_4}|$/GeV  & 255.5   &   104.1   &   509.6   &   169.7  \\\hline
$|m_{\tilde{\chi}^0_5}|$/GeV  & 255.8   &   104.9   &   509.7   &   199.1  \\\hline
$|m_{\tilde{\chi}^0_6}|$/GeV  & 256.0   &   105.2   &   509.8   &   199.4  \\\hline\hline
$|m_{\tilde{\chi}^\pm_1}|$/GeV& 254.6   &   101.8   &   509.1   &   169.7  \\\hline
$|m_{\tilde{\chi}^\pm_2}|$/GeV& 254.6   &   101.8   &   509.1   &   169.7  \\\hline\hline
$|R_{Z11}|$                   & 0.0036  &   0.0212  &   0.00090 &   $1.5\cdot 10^{-7}$ \\\hline
$|R_{Z12}|$                   & 0.0046  &   0.0271  &   0.00116 &   $1.7\cdot 10^{-4}$ \\\hline
$|R_{Z22}|$                   & 0.0018  &   0.0103  &   0.00045 &   0.106   \\\hline\hline
$X^{h_1}_{22}$                & 0.0044  &   0.0106  &   0.0022  &   0.00094 \\\hline
$\mathrm{Br}(h\rightarrow \tilde{\chi}^0_2 \tilde{\chi}^0_2)$& 4.7\%   & 21.9\%              & 1.23\% & 0.22\% \\\hline
$\mathrm{Br}(h\rightarrow b\bar{b})$                         & 56.6\%  & 46.4\%              & 58.7\% & 59.3\% \\\hline
$\Gamma(h\rightarrow \tilde{\chi}^0_2 \tilde{\chi}^0_2)$/MeV & 0.194   & 1.106               & 0.049  & 0.0088 \\\hline
$\Gamma^{tot}$/MeV                                           & 4.15    & 5.059               & 4.002  & 3.962 \\\hline
\end{tabular}
\caption{Benchmark scenarios for $m_{h_1}\approx 125\,\mbox{GeV}$. The branching ratios and decay widths of the
lightest Higgs boson, the masses of Inert neutralinos and charginos as well as the couplings of
$\tilde{H}^0_1$ and $\tilde{H}^0_2$ are calculated for $s=12000\,\mbox{GeV}$, $\lambda=0.6$, $\tan\beta=1.5$,
$m_{H^{\pm}}\simeq m_{A}\simeq m_{h_3}\simeq 9497\,\mbox{GeV}$, $m_{h_2}\simeq M_{Z'}\simeq 4450\,\mbox{GeV}$,
$m_Q=m_U=M_S=4000\,\mbox{GeV}$ and $X_t=\sqrt{6} M_S$.}
\end{table}

\end{document}